\documentclass{article}

\usepackage{arxiv}

\usepackage[utf8]{inputenc} 
\usepackage[T1]{fontenc}    
\usepackage{hyperref}       
\usepackage{url}            
\usepackage{booktabs}       
\usepackage{amsfonts}       
\usepackage{nicefrac}       
\usepackage{microtype}      
\usepackage{graphicx}
\usepackage{float}
\usepackage[frozencache,cachedir=.]{minted}
\usepackage{natbib}

\hypersetup{colorlinks=false}

\title{Including Physics in Deep Learning – An example from 4D seismic pressure saturation inversion}

\author{
Jesper S{\"o}ren Dramsch \\
Danish Hydrocarbon Research and Technology Centre \\
Technical University of Denmark \\
Building 375, 2800 Kongens Lyngby \\
Denmark \\
  \texttt{jesper@dramsch.net} \\
   \And
Gustavo Corte \\
Institute of Petroleum Engineering \\
Heriot-Watt University \\
Edinburgh, EH14 4AS \\
United Kingdom \\
  \texttt{gac2@hw.ac.uk} \\
  \AND
Hamed Amini \\
Institute of Petroleum Engineering \\
Heriot-Watt University \\
Edinburgh, EH14 4AS \\
United Kingdom \\
\texttt{h.amini@hw.ac.uk}
\And
Colin MacBeth \\
Institute of Petroleum Engineering \\
Heriot-Watt University \\
Edinburgh, EH14 4AS \\
United Kingdom \\
\texttt{c.macbeth@hw.ac.uk} \\
\AND
Mikael L{\"u}thje \\
DHRTC \\
Technical University of Denmark \\
Building 375, 2800 Kongens Lyngby \\
Denmark \\
\texttt{mikael@dtu.dk} \\
}

\begin{document}
\maketitle

\begin{abstract}
Geoscience data often have to rely on strong priors in the face of uncertainty. Additionally, we often try to detect or model anomalous sparse data that can appear as an outlier in machine learning models. These are classic examples of imbalanced learning. Approaching these problems can benefit from including prior information from physics models or transforming data to a beneficial domain.

We show an example of including physical information in the architecture of a neural network as prior information. We go on to present noise injection at training time to successfully transfer the network from synthetic data to field data.
\end{abstract}


\section{Introduction}

Physics in machine learning often relies on transformations of data to beneficial domains and simulating additional data. \citet{karpatne2017physics} show a physics-guided approach to model lake temperatures with neural networks. \citet{schutt2017quantum} use deep neural networks to model molecule energies and \citet{de2017learning} employ a special architecture to capture scatter patterns in high-energy physics. When building deep learning pipelines, we can make informed choices in data modeling, but also build neural networks to maximize information gain on the available data. \citet{ulyanov2018deep} has shown that the network architecture itself can be used as prior in machine learning. These approaches translate well to geoscience, where strong priors are often necessary to inform decisions.

Deep learning has revolutionized machine learning by replacing the feature generation and augmentation step by learned internal representations of features that maximize information gain. On image data analysis of these neural network filters have shown close relations to edge filters and color separators \citep{grun2016taxonomy}. \citet{dramsch2018deep} have shown that these filters translate well to seismic data. However, classic feed-forward neural networks do not have the benefit of learning filters. However, these neural networks benefit from recent improvements for regularization \citep{ioffe2015batch}, non-saturating and non-vanishing gradients \citep{he2015delving}, and training on GPUs.

Neural networks for inversion of seismic data have a long history \citep{roeth321}. In \cite{dramsc2019} we show the application of a deep multi-layer perceptron for map-based 4D seismic pressure saturation inversion. In this work we show the information gain of feed-forward multi-layer perceptron neural networks by including an explicit calculation of the AVO gradient within the network architecture. It's exemplary for including domain knowledge as a prior in machine learning.

\section{Method}
We build a deep feed-forward network to invert seismic amplitude maps for pressure and saturation changes. We use the high-level Python framework \texttt{keras} with a \texttt{tensorflow} backend. The neural network was trained on synthetic data, to subsequently predict field data. The network takes the seismic input samplewise with near, mid, and far stacks, and pore volume. We inject 20\% Gaussian noise to model the noisier field data directly after the input layer. This is fed to a custom layer that calculates the PP AVO gradient between far-mid, mid-near, and far-near. The main components are as follows:

\subsection{Gaussian noise injection}
The synthetic model is noise-free. While we get good results on the training data and the modelled test data, the network does not transfer well to noisy field data. Although the 4D NRMS is very low in the data set, the sample-wise fluctuations in the field seismic differ significantly from the synthetic data. We apply additive Gaussian noise with $\sigma = .02$ to the seismic inputs separately to simulate independent fluctuations of the seismic maps. This significantly decreases the training and validation performance on noise free synthetic data. On field data, however, this enables good transfer of the neural network.
\begin{minted}{python}
noisy_input = GaussianNoise(0.02)(input_data)
\end{minted}

\subsection{Explicit AVO gradient calculation}
The Schiehallion field is a good example of imbalanced learning. We have many samples of pressure changes $\Delta P$, a good selection of water saturation changes $\Delta S_w$, and very few gas saturation changes $\Delta S_g$. Yet, the changes in gas saturation $\Delta S_g$ produce the strongest changes in seismic P wave amplitudes. Statistically, these can easily be regarded as outliers, and therefore, possibly disregarded by the neural network. From decades of seismic analysis, we know that the AVO gradient is very good for pressure saturation separation. We implement an explicit calculation of AVO gradients in the network.
\begin{equation}
    G = \frac{A_{\Theta_1} - A_{\Theta_0}}{x_{\Theta_1} - x_{\Theta_0}},
\end{equation}
where $G$ is the PP AVO gradient, $A$ is the seismic P wave amplitude, $x$ is the offset, and $\Theta$ is the angle.
\begin{minted}{python}
mid_near = Lambda(
    lambda inputs: (inputs[0] - inputs[1]) / (10)
)([noisy_mid, noisy_near])

far_mid = Lambda(
    lambda inputs: (inputs[0] - inputs[1]) / (10)
)([noisy_far, noisy_mid])

far_near = Lambda(
    lambda inputs: (inputs[0] - inputs[1]) / (20)
)([noisy_far, noisy_near])
\end{minted}

\subsection{Encoder-decoder architecture}
Subsequently, the four input maps and the three gradient maps are concatenated and fed to an encoder architecture that condenses the information to an embedding layer $z$. This layer learns a collection of Gaussian distributions to represent the noisy input data The decoder samples this variational embedding layer to calculate the pressure change $\Delta P$, change in water saturation $\Delta S_w$, and gas saturation $\Delta S_g$. 

The full architecture is of the encoder-decoder class. The encoder reduces the number of parameters with each subsequent layer. This forces the network to learn a lossy compression of the input data as $z$-vector. The decoder increases the number of nodes per layer toward the output. The network therefore learns to correlate the low resolution representation with the desired output. 
\begin{figure}[ht]
    \centering
    \includegraphics[width=1.1\textwidth]{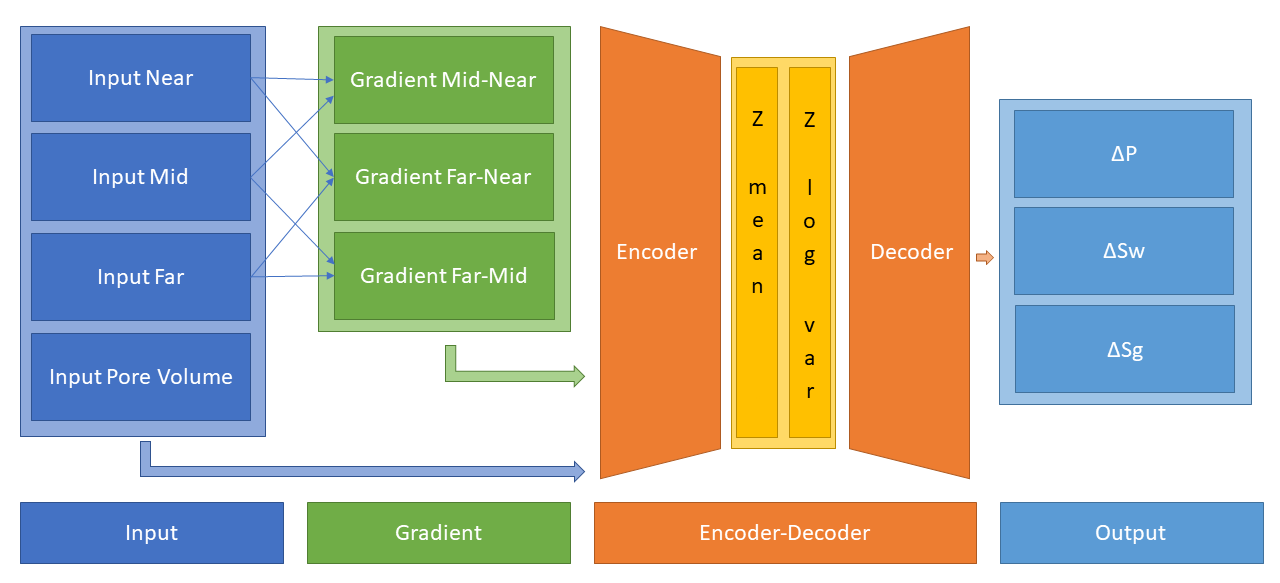}
    \caption{\vspace*{-12pt}Full Architecture from \citet{dramsc2019}.}\vspace*{-12pt}
    \label{fig:avonet}
\end{figure}

\subsection{Variational Z Vector}
The inversion of noisy input benefits from a variational representation of compressed z-vector. The networks learns Gaussian distributions in the embedding layer. Therefore, we have to apply the reparametrization trick outlined in \citet{kingma2013auto} to circumvent the sampling process cannot be learned by gradient descent. We use the implementation in \citet{chollet2015keras} for variational autoencoders.

\section{Results}
\begin{figure}[H]
    \centering
    \includegraphics[width=\textwidth]{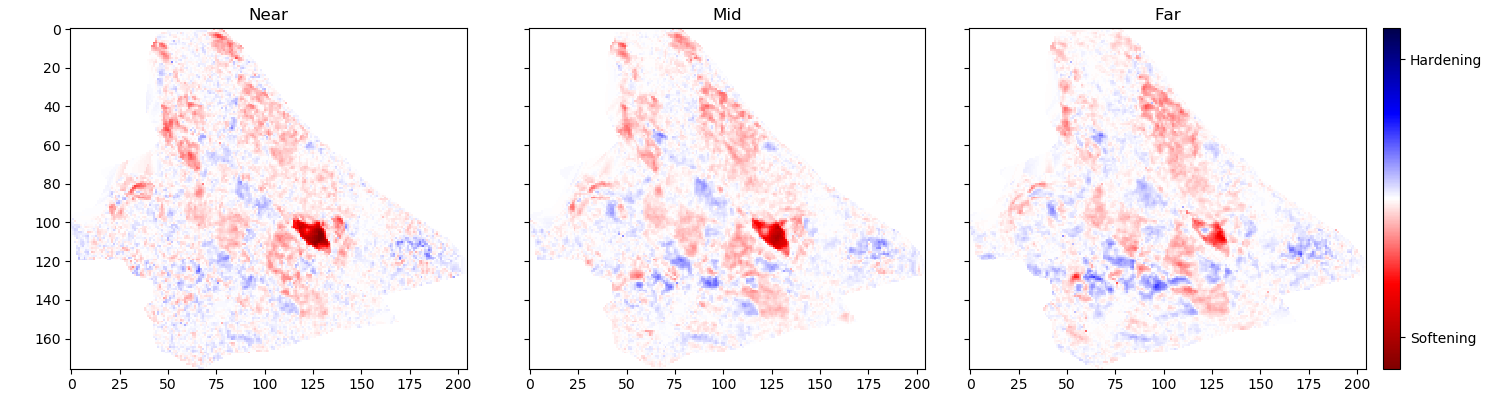}
    \caption{\vspace*{-12pt}Schiehallion 2004 Timestep Seismic data, pore volume and sim2seis results.}\vspace*{-12pt}
    \label{fig:input}
\end{figure}

\begin{figure}[H]
    \centering
    \includegraphics[width=\textwidth]{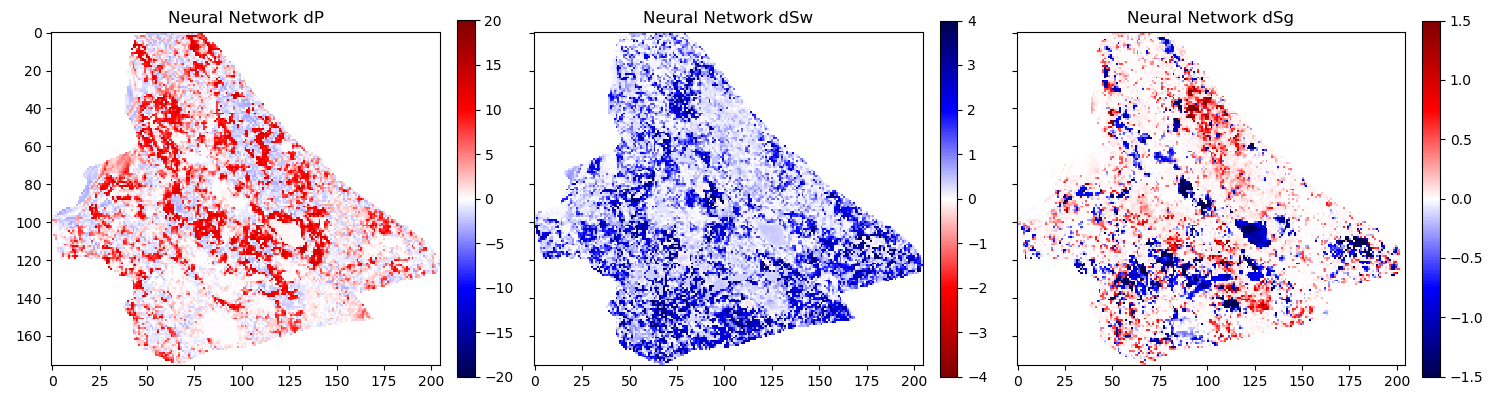}
    \caption{\vspace*{-12pt}Variational Encoder Decoder Architecture Inversion}\vspace*{-12pt}
    \label{fig:vae}
\end{figure}

In figure~\ref{fig:input} we show the 2004 time step of the Schiehallion 4D. Figure~\ref{fig:vae} contains the inversion result using the variational encoder decoder architecture. Some coherency in the maps can be seen, but each map is very noisy and the gas saturation map contains many data points that indicate gas desaturation, which cannot be confirmed by production data.

\begin{figure}[H]
    \centering
    \includegraphics[width=\textwidth]{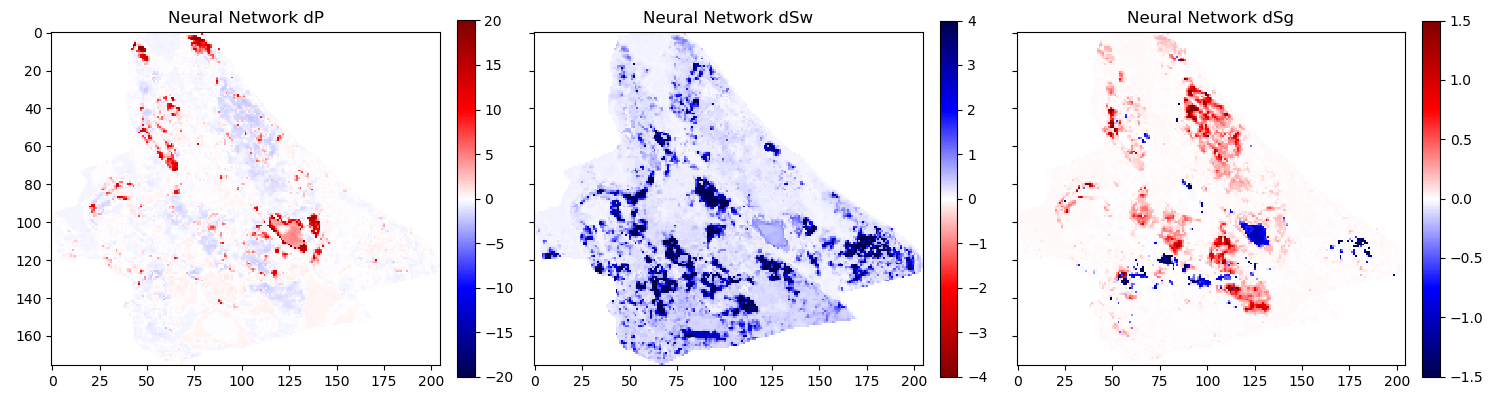}
    \caption{\vspace*{-12pt}Noiseinjected AVO-Gradient Variational Encoder Decoder Architecture Inversion}\vspace*{-12pt}
    \label{fig:noisegradvae}
\end{figure}

When we add the gradient, we can clean up some of the misfit in the gas saturation maps $\Delta S_g$. Particularly, the event with the strongest softening in the amplitude maps, is partially reassigned to the pressure map $\Delta P$. However, the inversion process is still very prone to noise. In figure~\ref{fig:noisegradvae}, we show the inversion results of a AVO-gradient neural network with a noise injection at training of $\sigma = .02$. The inversion maps are very coherent. Noise injection without gradient calculation does not give adequate results.
 
\begin{figure}[H]
    \centering
    \includegraphics[width=\textwidth]{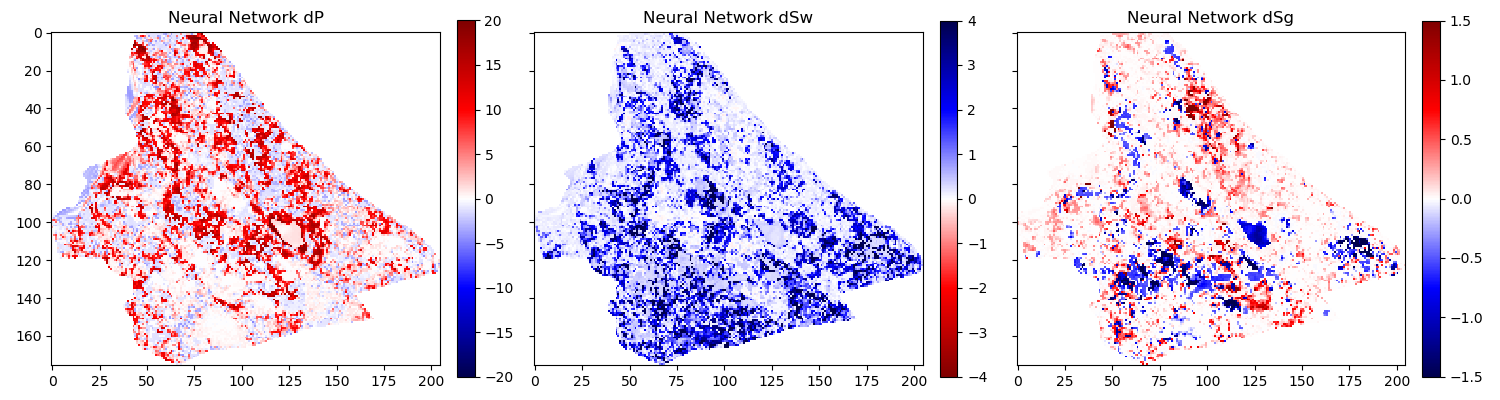}
    \caption{\vspace*{-12pt}AVO-Gradient Variational Encoder Decoder Architecture Inversion}\vspace*{-12pt}
    \label{fig:gradvae}
\end{figure}

\section{Conclusions}

We have shown a neural network architecture that incorporates physical domain knowledge to enable transfer from synthetic to field data. The final inversion result has very good coherency, despite the network not having any spatial context. While further investigation is necessary, this indicates that useful information has been learned. This is one example, where bias can be intentionally introduced into the network architecture to include physics into machine learning.

\section{Acknowledgements}
The research leading to these results has received funding from the Danish Hydrocarbon Research and Technology Centre under the Advanced Water Flooding program.  We thank the sponsors of the Edinburgh Time-Lapse Project, Phase VII (AkerBP, BP, CGG, Chevron, ConocoPhillips, ENI, Equinor, ExxonMobil, Halliburton, Nexen, Norsar, OMV, Petrobras, Shell, Taqa, and Woodside) for supporting this research. The Brazilian governmental research-funding agency CNPq. We are also grateful to Linda Hodgson and Ross Walder for important discussions on the field and dataset.

\bibliographystyle{unsrtnat}  
\bibliography{preprint}  

\end{document}